\documentclass[floatfix,prb,twocolumn,showpacs,amsmath,amssymb]{revtex4}
\usepackage[dvips]{graphicx}
\usepackage{latexsym}
\usepackage{amssymb}
\usepackage{amsmath}
\usepackage{amsbsy}
\newcommand{\Brace}[2]{\genfrac{\{}{\}}{0pt}{}{#1}{#2}}

\def\a{\alpha}
\def\afs{\alpha_{\rm fs}}
\def\b{\beta}
\def\g{\gamma}
\def\d{\delta}

\def\D{\Delta}
\def \e{\varepsilon}

\def\h{\hbar}
\def \oo {\omega}
\def\OO {\Omega}
\def\one{\openone}

\def\Tr{\mbox{tr}\,}

\def\S{{\mathcal S}}

\def\l{\lambda}

\def\DD{\partial}%  PARTIAL DERIVATIVE

\def\dwell{\tau_{\rm d}}

\def\RH{R_H}
\def\RHT{R_H^{\rm T}}

\def\A{{\mathcal A}}
\def\V{{\cal V}}
\def\la {\langle}
\def\ra {\rangle}

\def\RC {\tau_{\rm RC}}

\def\ie{{\it i.e.}}

\begin{document}
%\linenumbers
\title{Mesoscopic magnetoelectric effect in chaotic quantum dots}
\date{\today}
\author{M.~L.~Polianski}
\affiliation{ Niels Bohr Institute,
          NBIA,
          Blegdamsvej 17,
          DK-2100 Copenhagen
          Denmark}
\begin{abstract}
The magnitude of the inverse Faraday effect (IFE), a static
magnetization due to an ac electric field, can be strongly increased
in a mesoscopic sample, sensitive to time-reversal symmetry (TRS)
breaking. Random rectification of ac voltages leads to a
magnetization flux, which can be detected by an asymmetry of Hall
resistances in a multi-terminal setup. In the absence of applied
magnetic field through a chaotic quantum dot the IFE scale,
quadratic in voltage, is found as an analytic function of the ac
frequency, screening, and coupling to the contacts and floating
probes, and numerically it does not show any effect of spin-orbit
interaction. Our results qualitatively agree with a recent
experiment on TRS-breaking in a six-terminal Hall cross.
\end{abstract}
%73.63.Kv Quantum dots
%73.23.-b Electronic transport in mesoscopic systems

%75.80.+q Magnetomechanical and magnetoelectric effects, magnetostriction

%73.63.-b Electronic transport in nanoscale materials and structures
%(see also 73.23._b Electronic transport in mesoscopic systems)

%74.40.+k Fluctuations (noise, chaos, nonequilibrium
%superconductivity, localization, etc.)

%78.20.Ls Magnetooptical effects

 \maketitle
In pursuit of effects that combine spin and  charge, great attention
is paid to mutual effects of magnetic and electric fields. Their
magnetoelectric manipulation is interesting not only scientifically,
but also for possible applications.\,\cite{Fiebig_Revival:2005}
Obviously, any magnetooptic or magnetoelectric effect has its
counterpart. For example, a ferromagnet polarizes the spins of
electrons and conversely, a current can exert a torque on a
magnetization vector and switch a magnetic domain in random access
memory.\,\cite{Maekawa:concepts} Similarly, the Faraday effect, a
rotation of the polarization of light by a magnetic field, has its
inverse: a medium is magnetized by a beam of circularly polarized
radiation.\,\cite{Landau:8} We consider the unusual properties of
this magnetoelectric effect in small non-magnetic samples.

Classically, if a medium has a spatial inversion or time-reversal
symmetry (TRS), the expansion of its free energy $F$ does not have
terms linear in $E$ or $H$, respectively. Magnetoelectric effects
appear only due to mixing of $H$ and $E$, either as an $H_i E_j$
term if both symmetries are broken in equilibrium (magnetized
anisotropic medium) or in a higher order "$HEE$
term"\,\cite{Schmid:1994} $H_i E_j E_k^{(*)}$ if a strong electric
field breaks TRS. In particular, this $HEE$ term leads to
magnetization $M=\DD F/\DD H$, detectable in nonlinear effects like
second-harmonic generation or rectification\,\cite{Fiebig_2ndHG:05}.
The latter, known as inverse Faraday effect (IFE), is a static
magnetization induced by perturbations at the frequency $\oo$.
\,\cite{Landau:8,hertel:2005} An electrical ac voltage, $V_\oo$,
generated across a non-absorbing diffusive medium with mean-free
time $\tau$ induces an asymmetry of the dielectric tensor. Since it
is also linear in magnetic field, the classical IFE magnetization
flux, $\varphi_{\rm cl}$, can be estimated,\,\cite{Landau:8}
\begin{eqnarray}\label{eq:IFEcl}
\varphi_{\rm cl}\sim \afs\Phi_0 (eV_\oo)^2  m v_{\rm F}^3/(\h\oo)^3
c,\,\oo\tau\gg 1,
\end{eqnarray}
where $\afs$ is the fine structure constant, $\afs\approx 1/137$,
and $\Phi_0\equiv h/e$ is flux quantum. Importantly, this estimate
could as well be obtained from the Joule heating and the asymmetry
of the conductivity tensor. Since Eq.\,(\ref{eq:IFEcl}) contains
small $v_{\rm F}/c$ and the fine-structure constant, one naturally
asks: can we enhance the magnetic response to the $E$-field, which
eventually breaks TRS?

To show that it is possible to exceed Eq.\,(\ref{eq:IFEcl}), we
propose to use the sensitivity of electronic transport through a
mesoscopic sample to the broken TRS
(Ref\,\onlinecite{Beenakker:1991}) as a detector of magnetic flux.
Indeed, disorder inevitably breaks a spatial symmetry in such a
sample, and mesoscopic (sample-to-sample) fluctuations of transport
occur on a flux scale $\Phi\lesssim\Phi_0$. TRS breaking could be
induced either by a flux $\Phi$ of applied magnetic field, or by an
IFE flux $\varphi$ created by additional ac perturbations. At
$\Phi=0$, an indirect transport detection of $\varphi$ becomes
possible in a {\it multi-terminal}
 setup with a separate pair of current- and voltage-probes\,\cite{vanderpauw:1958} measuring the Hall (or non-local)
resistance
 $\RH$. Previously, Edelstein considered $\RH\propto V_\oo^2$ as a
 signature of IFE in a noncentrosymmetric diffusive two-dimensional material
 with the mirror symmetry broken by spin-orbit interaction (SOI).
 \,\cite{Edelstein:2005} Due to large spin-orbit scattering
 time, $\tau_{\rm so}\gg\tau\gg 1/\oo$, this effect is small,
 $\varphi_{\rm so}\sim \Phi_0(eV_\oo/\e_{\rm F})^2(\h/\oo\e_{\rm F}
 \tau_{\rm so}^2)$. In contrast, here we consider a large fluctuational
 effect.

Importantly, $\RH\neq 0$ even in small chaotic samples at zero
field, but its random response to $\Phi$ can be used as a
sample-specific gauge for the flux. B\"uttiker
showed\,\cite{Buttiker4terminal:1986} that the Onsager symmetry
relations hold in mesoscopics,
 $\RH(\Phi)=\RHT(-\Phi)$, where $\phantom {\!}^{\rm T}$ stands for the
 measurement with current- and voltage-probes exchanged; this prediction was
  confirmed in many linear transport
experiments.\,\cite{Benoit:1986,WashburnWebb:1992,Shepard:1992} Only
recently did Chepelianskii and Bouchiat show that these relations
are violated, when TRS is broken by additional ac
perturbations.\,\cite{Chepelianskii:2009} Since both applied $\Phi$
and induced $\varphi$ lead to the asymmetry $\RH\neq \RHT$, data in
$\RH(0)-\RHT(0)$ suggest a shift of zero magnetic flux, $\ie$ an
effective IFE flux $\varphi\neq 0$.

 In this Rapid Communication we develop a
theory that evaluates IFE in a multi-terminal chaotic quantum dot
subjected to external ac perturbations at the frequency $\oo$ in the
absence of applied magnetic field. First, we introduce a
(sample-specific) gauge for magnetic flux, using the response of
$\RH$ to small $\Phi$. Then we find the TRS-breaking in $\RH$ for
the perturbed dot and evaluate the scale of fluctuations of the
induced IFE flux $\varphi$ through its area:
\begin{eqnarray}\label{eq:IFEmeso}
\varphi\sim\pm \Phi_c (eV_\oo/\e)^2
%\mbox{min}\{1,\frac{\h}{\dwell T}\}
,\,eV_\oo\ll\e=\mbox{max }\{\h\oo,\h/\dwell,T\},
\end{eqnarray}
where $\Phi_c\lesssim\Phi_0$ is the flux that completely breaks
TRS,\,\cite{Beenakker:1997} $\dwell$ is a typical dwell time of
electrons in the dot, and $T$ is the temperature. Equation
(\ref{eq:IFEmeso}), the main result of our work, does not have
definite sign, which is typical for quantum effects. Similarly to
the equilibrium persistent current\,\cite{Levy:1990} or magnetic
response of quantum dots,\,\cite{Altshuler:93,Matveev:2000} it does
not contain $\afs$ or large $\e_{\rm F}$. Experimentally, a sample
can be magnetized not only by the flux $\Phi$ in equilibrium, but
also by ac voltages at $\Phi=0$. Specific details of the sample
rectify these voltages and lead to a random TRS breaking interpreted
as an IFE magnetization. We expect this quantum interference effect
to be measurable in any coherent sample of reduced dimensions. First
we explain the model and major steps in the derivation of
Eq.\,(\ref{eq:IFEmeso}). Then we discuss how IFE is affected by
screening, spin-orbit, and floating probes, and qualitatively
compare our predictions to experiment.\,\cite{Chepelianskii:2009}

\begin{figure}[b]
\centering\includegraphics[width=6.5cm]{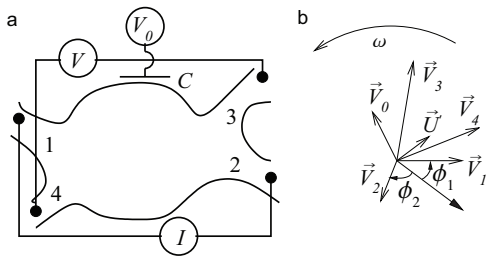} \caption{ (a) A
gated chaotic dot in a four-terminal set-up, where the current
through source-drain contacts 1 and 2 and the voltage drop between 3
and 4 yield $\RH=R_{12,34}$. IFE due to additional ac voltages
$V_\a\cos(\oo t+\phi_\a)$ induces a random magnetization flux
$\varphi$, detectable in $\RH-\RHT$; (b) Biases are specified by
$\vec V_\a=V_\a e^{i\phi_\a}$ in the complex plane. IFE depends on
the potential of the dot, $\vec U\approx\vec V_0$ (or $\vec U'$) for
weak (or strong) interactions. }\label{FIG:setup}
\end{figure}
 {\it Model.} We consider electronic transport through a multi-terminal
 $N$-channel chaotic quantum dot at the temperature $T$,
 see Fig.\,\ref{FIG:setup}(a), and for simplicity assume the dot
 to be circular with radius $L$. Chaos is set due to either the diffusive
 motion of electrons, $l=\tau v_{\rm F}\ll L$, or their random scatterings from
the boundaries, $L\ll l$, so that the ergodic time $\sim\mbox{max
}\{1, L/l\}L/v_{\rm F}$ needed to explore the dot is
short.\,\cite{Aleiner:2002,Beenakker:1997} Therefore, electronic
focusing and direct trajectories, present in some ballistic
structures,\,\cite{Chepelianskii:2009,Beenakker:1991} are absent.
Each of four contacts has $N_\a\gg 1$ ballistic orbital channels and
is characterized by $n_\a\equiv N_\a/N$. Additional coupling (not
necessarily ballistic) of floating
probes\,\cite{ButtikerPhysRevB:1986,Brouwer:1997} is discussed in
the end. ac voltages $V_\a\cos(\oo t+\phi_\a)$ applied at the same
frequency $\oo$, but generally out of phase, are for convenience
specified by the vectors $ \vec V_\a=V_\a e^{i\phi_\a}$ in a complex
plane [see Fig.\,\ref{FIG:setup}(b)]. A top gate 0 with capacitance
$C$ is biased by $\vec V_0$.

Screening in the dot is accounted for by the uniform time-dependent
potential $U$, and its higher spatial harmonics are suppressed due
to short ergodic time.\,\cite{Aleiner:2002} This potential is found
self-consistently from charge conservation and gauge
invariance.\,\cite{ButtikerPretre:1993PRL} The contributions leading
to Coulomb blockade are small in $1/N\ll 1$ and not taken into
account.\,\cite{BrouwerLamacraftFlensberg:2005} If $\D=2 \h^2/m L^2$
is the mean level spacing, the ratio $e^2/C\D$ defines the limits of
weak/strong interaction. Using the dwell time of electrons,
$\dwell=h/N\D$, we normalize the flux of applied magnetic field by
the crossover flux that completely breaks TRS, $\phi=\Phi/\Phi_c$,
where $\Phi_c/\Phi_0=
2L/\sqrt{\dwell v_{\rm F} l}%=\sqrt{2N}/\pi\eta
,l\ll L$ ($l\to \pi L/4$ if $L\ll l$).\,\cite{Beenakker:1997} We
assume that the dot is not perturbed internally, so that the
scattering is characterized by the energy-dependent scattering
matrix $\S(\e)$. Statistical averages over ensemble, denoted by
$\la...\ra$, are found to leading order in $1/N$ by diagrammatic
technique in
$\S(\e)$.\,\cite{BrouwerBeenakker:1996,PolianskiJPHYSA:2003} We
consider spinless electrons and normalize conductance/resistance
 by $\nu_s e^2/h$, where $\nu_s=2$ is spin-degeneracy, and
later compute the effect of SOI.\,\cite{Cremers:2003,Adam:2002}

The Hall resistance is measured as a linear response to additional
voltages $\V_\a$ applied at small frequency $\OO$ (experimentally,
the frequency of lock-in $\OO\lesssim 100$ Hz). If 1 and 2 are the
source and drain, and the voltages are measured at 3 and 4, one has
$I_1=-I_2=I$ and $I_3=I_4=0$ % giving , or
 defining $\RH\equiv
R_{12,34}=(\V_3-\V_4)/I_1$
(Ref.\,\onlinecite{Buttiker4terminal:1986}) (applied voltages and
current are related via
$R_{xx}=(\V_1-\V_2)/I_{1}\approx(1/n_1+1/n_2)/N$, whose fluctuations
might be neglected here). A transposition (exchange) of current- and
voltage-probes, while keeping the same ac perturbations at $\oo$ and
the applied magnetic flux, gives a different $\RHT\equiv
R_{34,12}=(\V_1-\V_2)/I_3$. Due to randomness of the voltage drop
$\V_{34(12)}$ across the dot, the {\it resistance} is more relevant
than the
 {\it conductance} considered %previously in open 2D
%diffusive samples
before.\,\cite{MaLee:1987}

{\it Derivation.} For a non-interacting dot one can use scattering
states \,\cite{Pedersen:1998} (or Jauho-Meir-Wingreen
formula\,\cite{Jauho:1994}) to express the current in a contact $\a$
as a function of perturbations in a probe $\b$. It responds not only
linearly to $\V_{\b\OO}$, but also to $V_{\b\oo}$, and for small
$eV_{\oo}/ \mbox{max }\{\h\oo,T,\h/\dwell\}\ll 1$ we expand current
to first orders in this small parameter. This yields a conductance
matrix $\tilde g_{\a\b}$ at $\OO\to 0$,
\begin{eqnarray}\label{eq:dIdV}
\tilde g_{\a\b}&=&\frac{\DD I_\a}{\DD \V_\b}\approx-\int d\e
\Tr\left[\one_\a\S(\e)\one_\b\S^\dag(\e)-\one_\a\one_\b\right]f'(\e)
\nonumber \\ && \mbox{} + \left(\frac{eV_{\b\oo}}{2}\right)^2\int
d\e\frac{
f(\e+h\oo)+f(\e-h\oo)-2f(\e)}{(\h\oo)^2}\nonumber \\
&&\times \DD_\e\Tr \one_\a\S(\e)\one_\b\S^\dag(\e) \equiv g_{\a\b}+
g''_{\a\b}(eV_{\b\oo}/2)^2.
\end{eqnarray}
While the first term is the usual dc conductance $g$, the second is
an out-of-equilibrium contribution. It is similar, but not
identical, to the photo-assisted current.\, \cite{Pedersen:1998}
However, Eq.\,(\ref{eq:dIdV}) does not satisfy gauge-invariance:
currents depend on $V_{\b\oo}^2$ and change if all voltages are
shifted by an arbitrary $\d V_\oo$. Therefore, we consider the
nearby gate and self-consistently find the internal potential of the
dot $U(t)$ due to linear screening: a potential $U_f$ on some
frequency $f$ is a linear combination of voltages at this frequency,
$\vec U_f=\sum_\b u_{\b,f}\vec V_{\b,f}$, where $u_{\b,f}$ are
complex characteristic potentials, which sum up to 1. This potential
depends on $\dwell$ and RC time $\RC=\dwell/(1+\nu_s e^2/C\D)$.
\,\cite{ButtikerPretre:1993PRL,PolianskiJPHYSA:2005}
 The overall voltage shift by $-\vec U$ reduces the problem to
 the non-interacting, and
now conductance, expressed via $F_\g\equiv e^2|\vec V_{\g}-\vec
U|^2/4$, reads as
\begin{eqnarray}\label{eq:tildeGdef}
\tilde g_{\a\b}&=&g_{\a\b} +\sum\nolimits _{\g=1}^4
g''_{\a\g}(\d_{\b\g}-u_{\b,0})F_\g.
\end{eqnarray}
Since the sum of $\tilde g_{\a\b}$ in Eq.\,(\ref{eq:tildeGdef}) over
$\a$ or $\b$ vanishes, this sample-specific degenerate $\tilde g$
satisfies charge conservation and gauge invariance, respectively:
unlike Eq.\,(\ref{eq:dIdV}), it depends on differences $\vec
V_j-\vec V_k$ unaffected by a voltage shift. Importantly, elements
of $\tilde g$ and their correlators depend both explicitly on the
static $u_{\b,0}$ and implicitly on dynamic $u_{\b,\oo}$. The matrix
$\tilde g_{2\times 2}\propto 1-\sigma_x$ is always
symmetric,\,\cite{footnote1} but the TRS breaking becomes noticeable
in a multi-terminal setup. Indeed, due to the symmetry to the matrix
transposition, $g(\phi)=g^T(-\phi)$, the matrices $g(0)$ and
$g''(0)$ are symmetric, but $\tilde g_{4\times 4}$ is not, However,
the measured result depends on the probe configuration.

Indeed, if some $i$-th probe does not draw any current, we can
eliminate the $i$-th row and column from $\tilde g$ and
simultaneously shift all voltages by $-\V_i$. We obtain $\RH$
inverting the remaining part $\tilde g_{3\times 3}$, and the same
method gives $R_{H}^{\rm T}$ in the other, transposed,
configuration. When all $V_{\g\oo}=0$, the substitute $\tilde g\to
g$ reproduces
\begin{eqnarray}\label{eq:Rlin}
 R_{H,\rm
dc}=\frac{g_{31}g_{42}-g_{32}g_{41}}{\mbox{det }g_{3\times 3}},\,
R_{H,\rm dc}^{\rm T}=\frac{g_{13}g_{24}-g_{14}g_{23}}{\mbox{det
}g_{3\times 3}},
\end{eqnarray}
and the symmetry of $g$ immediately gives the Onsager relation,
$R_{H,\rm dc}(\phi)=R_{H,\rm dc}^{\rm
T}(-\phi)$.\,\cite{Buttiker4terminal:1986} These Hall resistances,
always equal at $\phi=0$, start to differ at weak magnetic flux
$\phi\ll 1$, and to quantify their asymmetry we consider ${\cal
A}\equiv (R_{H,\rm dc}-R_{H,\rm dc}^{\rm T})/2\phi$. Its average
vanishes, $\la\A\,\ra=0$, and the Gaussian fluctuations are given by
% Var (\RH-\RH^T)/2=(D^2-C^2)/2*(1/N_1+1/N_2)(1/N_3+1/N_4)
%
\begin{eqnarray}\label{eq:Rdc}
\mbox{Var }{\cal A}\equiv \la{\cal A}^2\ra-\la{\cal
A}\ra^2=\frac{(n_1+n_2)(n_3+n_4)}{N^4 n_1 n_2n_3n_4}\int_0^\infty
{\cal I} d\tau,
\end{eqnarray}
where ${\cal I}\equiv (4/\dwell)e^{-\tau/\dwell}(\pi
T\tau/\h)^2\sinh^{-2} (\pi T\tau/\h)$. Compared to the classical
Hall effect in such dots, this sensitivity is $\sim (L/N\l_{\rm
F})^2\gg 1$ stronger, but depends on the widths of the probes. As
expected, when the dot is widely opened, $\Phi_c\to\Phi_0$ and
Eq.\,(\ref{eq:Rdc}) corresponds to the result of
Ref.\,\onlinecite{MaLee:1987} up to a numerical coefficient $\sim
1$.

With additional ac perturbations the Hall resistances
[Eq.\,(\ref{eq:Rlin})] gain IFE-corrections, and we find that $\d
R_\pm=\d \RH\pm\d \RHT$ are also normally distributed around zero,
with
\begin{eqnarray}\label{eq:SigmaDelta}
\mbox{Var }\d R_\pm &=&\frac{(n_1+n_2)(n_3+n_4)}{N^4 n_1
n_2n_3n_4}\int_0^\infty %\frac{{\cal I}(\tau)4(1-\cos\oo\tau)^2d\tau}{(\h\oo)^4}
{\cal I} d\tau \frac{(1-\cos\oo\tau)^2}{(\h\oo)^4}\nonumber \\
&&\times (X_{12}+X_{34}\pm 2e^{-4\phi^2\tau/\dwell}Y_{12}Y_{34}),\\
\Brace{X}{Y}_{\a\b} &=&\frac{n_\a
n_\b}{n_\a+n_\b}\Brace{F_\a^2/n_\a+F_\b^2/n_\b-(F_\a-F_\b)^2}
{F_\a/n_\a+F_\b/n_\b}\nonumber,
\end{eqnarray}
 Interestingly, the aforementioned
dependence of $\tilde g$ and its fluctuations on $u_{\b,0}$ vanishes
from Eq.\,(\ref{eq:SigmaDelta}) due to the antisymmetry of
Eq.\,(\ref{eq:Rlin}). Therefore, the statistics of $\d R_\pm$ is
unaffected by the static characteristic potentials.

The magnetization IFE flux can now be evaluated using the TRS
breaking $\d R_-$ at $\phi=0$, $\varphi=\d R_-/2\A$. To leading
order, we can take $\A$ and $\d R_-$ uncorrelated and using
Eqs.\,(\ref{eq:Rdc}) and (\ref{eq:SigmaDelta}) find a Lorentzian
mesoscopic distribution
$P(\varphi)=\sigma/[\pi(\varphi^2+\sigma^2)]$ with
\begin{eqnarray}\label{eq:fife}
\sigma^2=\frac{\int d\tau {\cal I}(1-\cos\oo\tau)^2 }{4(\h\oo)^4\int
d\tau{\cal I}}[X_{12}+X_{34}- 2Y_{12}Y_{34}]\Phi_c^2.
\end{eqnarray}
The odd moments of $P(\varphi)$ vanish and the even ones diverge,
but $\varphi$ is solely defined by $\sigma$,
$\la|\varphi/\sigma|^{\pm 1/2}\ra=\sqrt{2}$. In case one energy $\e$
among the energy scales $\h\oo,\h/\dwell,T$ is large compared to the
other two, the ratio in Eq.\,(\ref{eq:fife}) equals to $1/\e^4$
multiplied by 3/8, 3/2 or 1/112 respectively, which results in Eq.\,
(\ref{eq:IFEmeso}). The $\sigma^2$, a positive-semidefinite form of
$F_\g$, can be created by a single $\vec V$ in some contact and
 vanishes only in a degenerate situation when all
 $F_{\g}$ are the same. One such example is when all voltages in the contacts
 have the same magnitude and phase and a gate-voltage
 $\vec V_0\neq \vec V_\g$ cannot induce IFE,
see discussion of experiment. This situation is similar to a
linearly polarized $\vec E$ not being able to induce the classical
IFE, $\vec M\propto [\vec E\times\vec E^*]=0$. Unless this uniform
regime is chosen to diminish $\varphi$, Eq.\,(\ref{eq:IFEmeso})
remains a good order-of-magnitude estimate of IFE. As a function of
frequency, $\sigma$ in Eq.\,(\ref{eq:fife}) is modified by
screening: in a strongly interacting dot, $\RC\ll\dwell$, in the
high-frequency limit, $\oo\RC\gg 1$, the
 capacitor is short-cut compared to the contact resistances and
 $\vec U\approx\vec V_0$. As a result, IFE can occasionally become
stronger than $1/\oo^2$ due to the increased magnitude of $|\vec
V_\g-\vec U|$ for the particular configuration of voltages [see
Fig.\,\ref{FIG:setup}(b)].
\begin{figure}[t]
\centering\includegraphics[width=8.5cm]{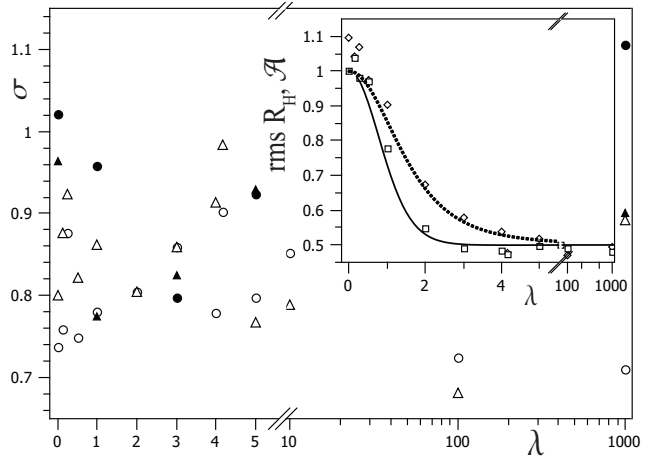} \caption{ The
width $\sigma$ ($\circ$) of the IFE Lorentzian distribution and
$\sigma_{\rm m}$ ($\triangle$) as functions of spin-orbit strength
$\l$, normalized by $\sigma$ of Eq.\,(\ref{eq:fife}); (inset) rms of
$\RH$ ($\diamond$) [$\A$ ($\square$)], normalized by $8[16]/N^2$,
compared with theoretical dotted [solid] curves. Empty (filled)
symbols correspond to $N=16(32)$.} \label{FIG:numerics}
\end{figure}

Until now, spin-orbit was neglected in the derivation of
Eq.\,(\ref{eq:fife}). To take it into account we construct a
scattering matrix $\S(\e)$, which depends on magnetic field and
SOI-strength, compute $g$ and $\tilde g$, and find IFE flux
$\varphi=\d R_-/2\A$. For illustration, we take a simplified
model\,\cite{Adam:2002} of SOI using a parameter $\l\sim
\h/\Delta\tau_{\rm so}$ (Ref.\,\onlinecite{Cremers:2003}) for the
dot's random Hamiltonian $\cal H$: in the limits $\l=0(\infty)$ it
belongs to the Gaussian orthogonal (symplectic) Ensemble GOE, $\b=1$
(GSE, $\b=4$). In the crossover region the $2M\times 2M$ Hamiltonian
with a fixed $\D$ is represented by ${\cal H}=[\sqrt{M}{\cal
H}_1+(\l/2) {\cal H}_4]/\sqrt{M+\l^2}$. Our numerics are done at
$M=25N$ to ensure $M\gg N$, for $T=0$, $\oo\ll\h/\dwell$, $N_i=N/4$,
and only $V_1\neq 0$. To find the IFE scale we fit $\sim 1200$
sample statistics to a shifted Lorentzian, and the result is
presented in Fig.\,\ref{FIG:numerics} together with $\sigma$ given
by moments, $\sigma_{\rm
m}=\la|\varphi|^{1/2}\ra/\la|\varphi|^{-1/2}\ra$. Deviations of
$\sigma(\l)$ from Eq.\,(\ref{eq:fife}) are attributed to a
relatively small number of channels and appear to be nonsystematic.
The inset in Fig.\,\ref{FIG:numerics} shows rms of $\RH$ and $\cal
A$ from the fits to normal distributions in very good agreement with
our predictions $0.5(1+3/(1+4\l^2/N)^2)^{1/2}$ and
$0.5(1+(3-4\l^2/N)/(1+4\l^2/N)^4)^{1/2}$, respectively. The lack of
any systematic trend of the available statistics in the main plot
(compared to the inset) suggests that the mesoscopic IFE is
unaffected by SOI, in contrast with the weak IFE existing only due
to SOI.\,\cite{Edelstein:2005}

The generalization from a 4-contact setup to $M$ contacts includes
arbitrarily coupled floating contacts, which are typical in a Hall
measurement. The voltage in a floating probe is adjusted to allow
exchange of electrons with the dot, but not to draw any current. As
a result, our analytical results become renormalized using
$N=\sum_{i=1}^4 N_i+N_f$. Not necessarily integer, $N_f$ is defined
similarly to $N_i$ by the total dimensionless conductance $N_f$ of
all available floating contacts. In the dephasing probe
model\,\cite{ButtikerPhysRevB:1986} the inelastic scattering with
time $\tau_\phi$ adds $h/\Delta\tau_\phi$ fictitious channels to
$N_f$. This redefinition decreases $\dwell$ and $n_\a$ used in $\cal
I$ and Eqs.\,(\ref{eq:Rdc})--(\ref{eq:fife}).

{\it Discussion of experiment.}
Reference\,\onlinecite{Chepelianskii:2009} measured a TRS breaking
at $\phi=0$ in gated ballistic Hall samples, where each of 6 (=4+2
floating) contacts had $N_\a\sim 200$ channels. AC perturbations
were applied at $\oo\sim 10^6-10^{10}$Hz, either asymmetrically
(sample A) or uniformly (sample B). Corrections to $\RH$,
%(equilibrium contribution was subtracted from data)
 quadratic in small perturbation amplitude, were measured as
functions of $\oo$ in the sample A. They vanish for $\oo<\oo_0$ and
for $\oo>\oo_c$ become of the same order, and a single-parameter
pumping in a billiard was used to numerically reproduce
non-monotonic $\RH-\RHT$ as a function of $\oo$. On the contrast,
sample B showed no TRS breaking, $\d\RH\approx \d\RHT$.

Alternatively, data can be interpreted as mesoscopic fluctuations of
IFE in a chaotic sample. They are visible even at large $N$ due to
 vanishing classical effect. Unfortunately, data can not be
directly compared with RMT result (\ref{eq:SigmaDelta}) because the
Hall cross does allow direct trajectories resulting in much smaller
non-local voltages and at 0.3 K it is beyond the universal
regime.%, is much larger than even the Thouless energy of the closed
%sample, $\h/\erg\approx 50 \mbox{mK}$
 Experiment interprets $\oo_0$ as a frequency characteristic for
(unknown) capacitive coupling with the contacts, when ac voltages
become noticeable. In our setup the voltages are given and this
circuit effect is not accounted for. At low frequencies  $\d R_\pm$
vary $\propto (\oo\dwell)^2$, but the anti-symmetric component is
generally smaller by construction [cf. $\pm$ in Eq.
(\ref{eq:SigmaDelta})]. Beyond the threshold $\oo_c\sim 1/\dwell$
both fluctuate similarly with a typical period $\D\oo\sim 1/\dwell$.
 Indeed, in
the experiment $\d R_+$ is usually larger and $\oo_c\sim\D\oo$ is
consistent with $\dwell\approx$3 ps expected from a dot with the
sample area and $N\approx 1200$. For the uniformly perturbed sample
B with equal voltages $\vec
 V_{1...4}$, Eq. \,(\ref{eq:SigmaDelta}) indeed
 results in fluctuations of $\d R_+$ and $\d R_-=0$ observed in experiment,
 which does not enter a high frequency regime.
 However, it is desirable to perform measurement in a truly chaotic
sample without direct trajectories and at lower $T$ and $N$ to
increase quantum fluctuations, or measure IFE directly using
superconducting quantum interference device, similarly to persistent
current.\,\cite{Levy:1990}

{\it Conclusions.} We consider a time-reversal symmetry breaking by
external ac voltages. Rectified perturbations generate a static
magnetoelectric effect, a random magnetization of a mesoscopic
sample, similar to the classical Inverse Faraday Effect. Mesoscopic
fluctuations of the magnetization flux can be measured using
out-of-equilibrium transport in a multi-terminal quantum dot. We
estimate a typical flux, quadratic in voltages, as a function of
frequency, screening, coupling to reservoirs, and numerically find
that spin-orbit scattering has a very small effect on IFE. Our
results allow qualitative comparison with and explain most important
features of a recent experiment in TRS breaking.

{\it Acknowledgements.} Alexei Chepelianskii suggested this problem
to me, and I thank him, H\'el\`ene Bouchiat and Markus B\"uttiker
for discussions and comments.

%\bibliography{BIBLIO}

\begin{thebibliography}{32}
\bibitem[{foo()}]{footnote1}
\bibinfo{note}{IFE measured in a 2-terminal setup using the position of the
  weak localization peak does not survive mesoscopic averaging, and a
  sample-specific result is smeared out by UCF.}
\end{thebibliography}

%%%%%%%%%%%%%%%%%%%%%%%%%%%%%%%%%%%%%%%%%%%%

\end{document}